\documentclass[aps,prd,amssymb,showpacs,floatfix,preprint,superscriptaddress,nofootinbib]{revtex4-1}
\usepackage{amsmath,amsfonts,graphics,epsfig,amssymb}

\begin{document}
\title{%
Constraining scenarios of the soft/hard transition for the pion
electromagnetic form factor with expected data of 12-GeV Jefferson Lab
experiments and of the Electron--Ion Collider }

\author{S.V.~Troitsky}

\affiliation{Institute for Nuclear
Research of the Russian Academy of Sciences, 60th October Anniversary
Prospect 7a, Moscow 117312, Russia}

\email{st@ms2.inr.ac.ru}

\author{V.E.~Troitsky}

\affiliation{D.V.~Skobeltsyn Institute of Nuclear Physics,
M.V.~Lomonosov Moscow State University, Moscow 119991, Russia}

\email{troitsky@theory.sinp.msu.ru}

\pacs{13.40~Gp, 14.40~Be, 12.39~Ki, 11.10~Jj}

\date{December 31, 2014}

\begin{center}
\begin{abstract}
It has been shown previously \cite{PRD} that a
non-perturbative relativistic constituent-quark
model for the $\pi$-meson electromagnetic form factor
allows for a quantitative description of the soft/hard transition,
resulting in the correct Quantum-Chromodynamical asymptotics, including
normalization, from the low-energy data without further parameter tuning.
This happens universally whenever the constituent-quark mass is switched
off. The energy range where the transition happens is therefore determined
by the quark-mass running at intermediate energies and is not tightly
constrained theoretically. Here we consider possible ways to pin down this
energy range with coming experimental data. We demonstrate that expected
experimental uncertainties of the 12-GeV Jefferson-Lab data are larger
than the span of predictions of the model, so these data might be used for
testing the model but not for determination of the soft/hard transition
scale. Contrary, the projected Electron-Ion Collider will be capable of
pinning down the scale.
\end{abstract}
\end{center}
\maketitle


\section{Introduction}
\label{sec:intro}
Making connections between high-energy (hard) and low-energy (soft) models
of hadrons, that is between the Quantum Chromodynamics (QCD) and effective
theories of strong interactions working in the infrared limit, is one of
the major challenges of modern particle theory. The electromagnetic form
factor of the charged $\pi$ meson, $F_{\pi}$, represents a particularly
interesting observable in this context. On one hand, its high-energy
asymptotics is well defined within the perturbative
QCD~\cite{FarrarJackson, EfremovRadyushkinAs, LepageBrodskyAs}. On the
other hand, precise experimental data in the soft region allow to directly
trace the evolution of the observable with the momentum transfer, $Q^{2}$,
up to $Q^{2}\gtrsim 2$~GeV$^{2}$, that is close to the range where one
expects the hard behavior to start settling down.

The $F_{\pi}$ form factor is remarkable in one more aspect which we will
exploit here. It is probably the only observable for which a successful
low-energy theory exists which gives the correct QCD asymptotics
quantitatively without a dedicated parameter tuning~\cite{PRD}. The
soft/hard transition is governed by switching the constituent-quark mass
$M$ off in the model, and wherever the mass is switched off, the QCD
asymptotics settles down, provided the low-energy description works well.
Therefore, no additional parameters are required to be tuned to reproduce
quantitatively the QCD asymptotics. The model uses the $M(Q^{2})$
dependence as an input, and various ways of switching the mass off are
allowed, just because the asymptotics is universal and is determined by
the infrared, and not intermediate-scale, parameters. Therefore, the
model, in its present form, does not predict the energy scale at which the
soft/hard transition, that is switching $M(Q^{2})$ off, takes place; this
scale is to be determined either by a detailed model for $M(Q^{2})$ or
experimentally. Unfortunately, detailed models for the $M$ running
available in the literature contain a number of free parameters and can
hardly be used for this purpose. Here, we explore prospects for
experimental determination of this scale from future $F_{\pi}$
measurements.

The rest of the paper is organized as follows. In Sec.~\ref{sec:model}, we
briefly review the model for the pion form factor which gives the correct
QCD asymptotics starting from the low-energy physics. In
Sec.~\ref{sec:constraints}, we describe how the unknown ingredient of the
model, the dependence of the effective constituent-quark mass on $Q^{2}$,
is constrained. We proceed in Sec.~\ref{sec:results} with estimates of the
impact future data may have on these constraints, working within two
representative scenarios; then we briefly conclude in Sec.~\ref{sec:concl}.

\section{The model frameworks}
\label{sec:model}
We adopt a well-elaborated model for the pion form
factor~\cite{PRD, KrTr-EurPhysJ2001, KrTr-PRC2002, KrTr-PRC2003,
KrTr-PRC2009, EChAYa2009} originally developed as a low-energy theory
based on the
Poincar\'e invariant constituent-quark model. It exploits the instant form
of Relativistic Hamiltonian Dynamics (see e.g.\
Ref.~\cite{KeisterPolyzou}). The use of the Modified Impulse
Approximation~\cite{KrTr-PRC2002} provides for the full relativistic
invariance and eliminates certain drawbacks of the original instant form.
The pion form factor, $F_{\pi}(Q^{2})$, is given by rather cumbersome but
explicit expressions~\cite{KrTr-PRC2002} collected in Ref.~\cite{PRD},
which we do not quote here. The low-energy model has two free parameters,
the constituent-quark mass $M$ and the wave-function confinement scale $b$
(the actual choice of the wave function does not have any significant
effect on the result, see Refs.~\cite{KrTr-EurPhysJ2001, KrTr-PRC2009}).
These two parameters are tuned to reproduce correct experimental values of
the pion decay constant $f_{\pi}$ and of the pion charge radius. It is
remarkable that these parameters were fixed from the low-energy data
(actually in 1998, Ref.~\cite{KrTr-EurPhysJ2001}, by making use of the
$F_{\pi}(Q^{2})$ measurements at $Q^{2}\lesssim
0.26$~GeV$^{2}$~\cite{Amendolia}), so no room to tune them remained. The
predictions of Ref.~\cite{KrTr-EurPhysJ2001} have been subsequently
verified by new measurements of $F_{\pi}(Q^{2})$ up to $Q^{2}\simeq
2.5$~GeV$^{2}$, an order of magnitude higher, and are in excellent
agreement with all present-day data.

In parallel with this phenomenological success, the model has an
interesting, if not miraculous, theoretical advantage. It has been noted
in Ref.~\cite{KrTr-asymp} that the asymptotical behavior $Q^{2}
F_{\pi}(Q^{2}) \sim \mbox{const}$, predicted by QCD~\cite{Qcounting1,
Qcounting2}, is obtained in this model at $M\to 0$. This is however not a
full story: introducing explicit $M(Q^{2})$ dependence in such a way that
$M(0)$ is taken from the original model but $M(\infty)=0$, we obtained in
Ref.~\cite{PRD} the numerical coefficient of this asymptotics which
appeared to reproduce the QCD predictions quantitatively, without any
parameter tuning, for every possible way of switching the quark mass off.
The QCD asymptotics~\cite{FarrarJackson, EfremovRadyushkinAs,
LepageBrodskyAs} is
\begin{equation}
Q^{2} F_{\pi}(Q^{2}) \simeq 8 \pi \alpha_{s}^{(1)}(Q^{2}) f_{\pi}^{2},
\label{Eq:*}
\end{equation}
where $\alpha_{s}^{(1)}(Q^{2})$ is the one-loop QCD coupling constant
(extension to higher loops is not straightforward, see e.g.\
Refs.~\cite{Bakulev?, LepageBrodskyRev}). The right-hand side of
Eq.~(\ref{Eq:*}) is determined by two parameters, $f_{\pi}$ (determined in
the low-energy theory) and the QCD scale $\Lambda_{\rm QCD}$ (which is
related, though not explicitly, to the low-energy confinement parameter
$b$). The fact that, by fixing $f_\pi$ and the charge radius in the
low-energy theory, we immediately reproduce Eq.~(\ref{Eq:*})
quantitatively, is an important advantage of the model, not seen in other
approaches.

The idea of switching the constituent-quark mass $M$ off in order to
obtain the form-factor behaviour at high $Q^{2}$  was put forward in
Ref.~\cite{Kiss} in the frameworks of the light-front approach
\cite{KissLF}; however, Eq.~(\ref{Eq:*}) was not obtained there. In
Ref.~\cite{PRD}, we used a parametrization for $M(Q^{2})$ inspired by
Ref.~\cite{Kiss} but corrected for effects of a one-gluon exchange,
\begin{equation}
M(Q^{2})=M(\infty) + \left(M(0)-M(\infty) \right)
\frac{1+{\rm e}^{-\mu^{2}/\lambda^{2}}  }{1+{\rm
e}^{(Q^{2}-\mu^{2})/\lambda^{2}} } L(Q^{2}),
\label{Eq:5*}
\end{equation}
\begin{equation}
L(Q^{2})=\frac{1}{1+\log \frac{Q^2 +\mu^{2}}{\mu^2}}.
\label{Eq:5**}
\end{equation}
The soft/hard transition is therefore governed by two parameters, $\mu$
and $\lambda$, of the $M(Q^{2})$ function.
The $\mu$ and $\lambda$ parameters determine the position and the
steepness of the transition between $M(0)$ and $M(\infty)$. The boundary
value $M(\infty) \to 0$ while $M(0)=0.22$~GeV is fixed, like in previous
studies, to reproduce $f_{\pi}$ and the pion charge radius correctly. We
note that the parametrization (\ref{Eq:5*}), (\ref{Eq:5**}) describes well
the $M(Q^{2})$ dependence obtained, within certain assumptions about free
parameters, in complicated non-perturbative dynamical models~\cite{Posled,
nucl-th/0005015}, see Ref.~\cite{PRD} for details and illustrations.

To summarize this section, we have a predictive quantitative model for the
pion form factor whose low-energy parameters are fixed and determine the
correct high-energy asymptotics automatically, but the soft/hard
transition is parametrized by $M(Q^{2})$, that is by two parameters $(\mu,
\lambda)$. We turn now to constraining these parameters.

\section{Constraining $M(Q^{2})$}
\label{sec:constraints}
\begin{figure}
\centering
\includegraphics[width=0.9\columnwidth]{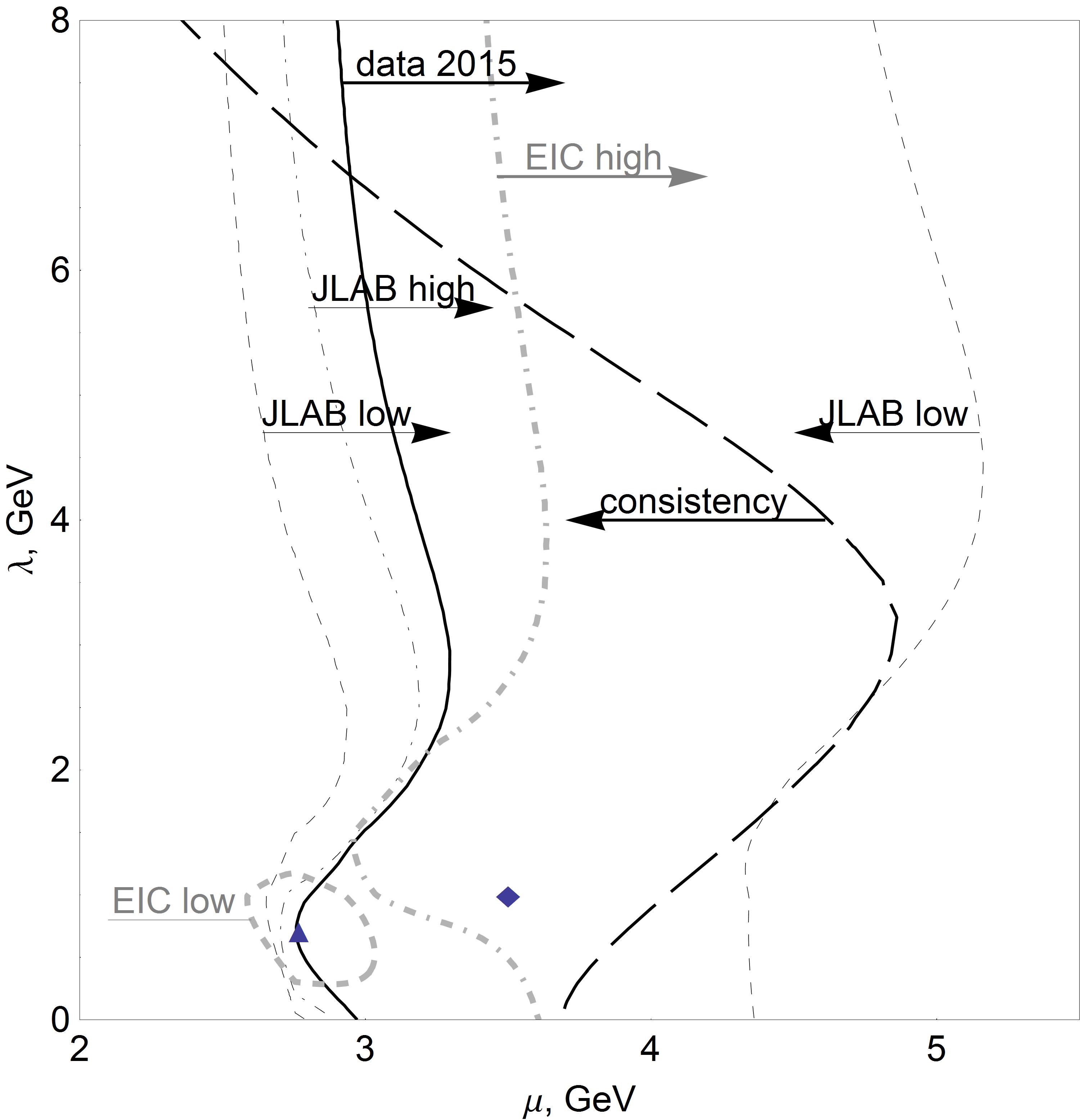}
\caption{
Allowed regions of the plane of parameters $\mu$ and $\lambda$ which
govern the quark-mass evolution, see text. Long-dashed line: the
consistency limit (region to the left of the line is allowed; relevant
for all cases). Full black line: the 95\% CL bound from the present data
(region to the right of the line is allowed). Other lines: 95\% CL
example bounds from future data for the ''low-scale'' (dashed; the allowed
region is bound by the lines) and ``high-scale'' (dot-dashed; the allowed
region is to the right from the lines) scenarios of the soft/hard
transition; thin lines assume 12-GeV JLab data, thick gray lines
assume EIC data. The values of $\mu$ and $\lambda$ assumed for
the ``low-scale'' and ``high-scale'' scenarios are shown by the triangle
and the diamond, respectively. }
\label{fig:mu-lambda}
\end{figure}

We start with theoretical constraints which are determined by the limits
of applicability of the model. While, technically, Eq.~(\ref{Eq:5*})
implies that $M$ is always decreasing for arbitrary $\mu$ and $\lambda$,
it appears that it cannot decrease too slow. Indeed, for each fixed
$Q^{2}$, there exists a value $M_{\rm max}(Q^{2})$ beyond which the form
factor, considered as a function of $M$, ceases to be monotonic. This, in
turn, implies that the form factor as a function of $Q^{2}$ may cease to
be monotonically decreasing. We calculate this $M_{\rm max}(Q^{2})$
numerically and determine, from the requirement $M(Q^{2})<M_{\rm
max}(Q^{2})$, the corresponding restriction on the parameters $(\mu,
\lambda)$. This bound (``the consistency limit''), applicable in any case,
is presented in Fig.~\ref{fig:mu-lambda}
as a
long-dashed line: the allowed region is to the left of this line. Note
that the requirement of perturbativity at
large $Q^{2}$ we used in Ref.~\cite{PRD} is qualitatively similar to this
bound.

The other kind of constraints come from experimental measurements of
$F_{\pi}$ at relatively high $Q^{2}$. As we have already pointed out, the
original model with $M={\rm const}$ predicted the form-factor values up to
$Q^{2}\sim 2.6$~GeV$^{2}$ with high accuracy, hence an early departure
from the constant-mass scenario might result in a disagreement with data.
The corresponding bound on $M(Q^{2})$ is obtained from the requirement of
agreement, at the 95\% confidence level (CL), of the corresponding
$F_{\pi}(Q^{2})$ function with the data points, tested
by
means of the ususal chi-square method.
For demonstration purposes, we also define the ``soft/hard
transition scale'' $Q^{2}_{\rm trans}$ as the value of $Q^{2}$ at which
the difference between the predicted $Q^{2} F_{\pi} (Q^{2})$ and
Eq.~(\ref{Eq:*}) is one half of its maximal value, that is the form factor
is half way from its nonperturbative values to the QCD asymptotics.

\begin{figure}
\centering
\includegraphics[width=0.67\columnwidth]{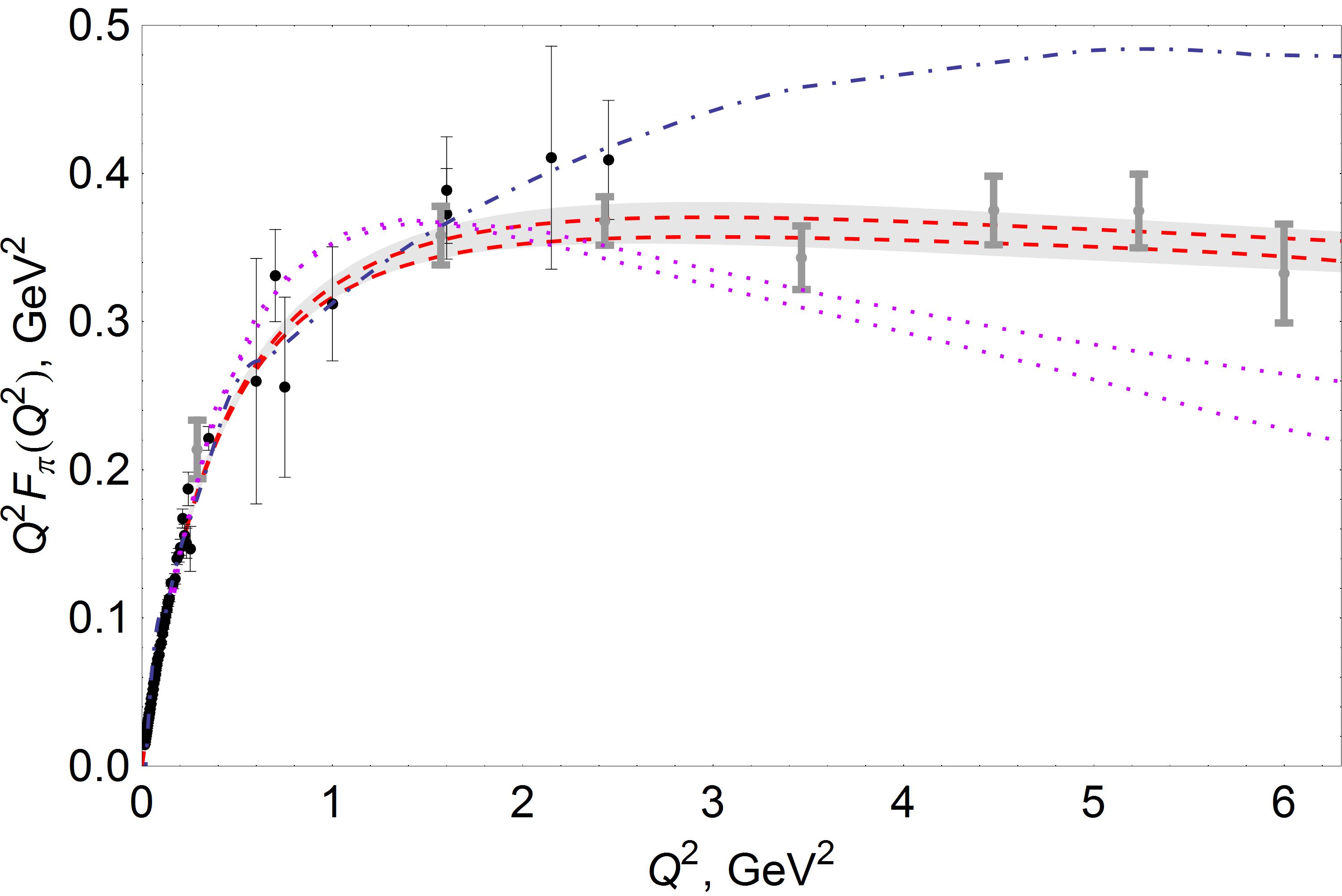}
\caption{
The range of predictions of our model for the charged pion form factor
allowed by the present data and model consistency for various scenarios of
quark-mass running (gray band) together with predictions of
Refs.~\cite{MarisRoberts} (area between two dotted lines) and
\cite{newTandy} (dash-dotted line). Two dashed lines represent two
particular scenarios for $M(Q^{2})$ corresponding to the ``low-scale'' and
``high-scale'' soft/hard transitions discussed in the text. Existing
experimental data points (see Ref.~\cite{PRD} for their description and
list of references) are shown by black dots with thin error bars.
A typical simulated example realization of expected 12-GeV JLab data,
corresponding to the ``high-scale'' scenario, is shown by gray dots with
thick gray error bars. It is evident that the JLab data would not help to
choose between the allowed scenarios of quark-mass running within our
model but would provide a good test of the model versus others.}
\label{fig:plot-6GeV2}
\end{figure}

By making use of all present-day data described in Ref.~\cite{exp-data},
we obtain constraints on $M(Q^{2})$ which are presented in
Fig.~\ref{fig:mu-lambda} in terms of $\mu$ and $\lambda$ (the full line;
the region to the right of the line is allowed). The corresponding range of
allowed $F_{\pi}(Q^{2})$ is shown as a gray band in
Fig.~\ref{fig:plot-6GeV2}
(see Ref.~\cite{PRD}  for more plots). In terms
of the soft/hard transition scale, this constraint is $Q^{2}_{\rm trans} >
8.5$~GeV$^{2}$.

Having determined the constraints on $M(Q^{2})$ from the present data, we
are ready to discuss prospective bounds from future experiments.

\section{Expected constraints from future data}
\label{sec:results}
Experimental prospects of the measurements of the pion form factor are
briefly summarized in Ref.~\cite{Horn2014}. They include the approved
E12-06-101 experiment at the upgraded Jefferson Laboratory (12-GeV JLab)
fascility and measurements at the projected Electron-Ion Collider (EIC).
More details may be found in the experimental proposal \cite{PR12-06-101}
for the 12-GeV JLab and in the talk \cite{EIC-talk} for EIC. Hereafter, we
will use the information about the $Q^{2}$ reach and projected error bars
of $F_{\pi}$ measurements at these fascilities given in
Refs.~\cite{PR12-06-101, EIC-talk} and reproduced in Ref.~\cite{Horn2014}
(for EIC, we assume the energy of the ion beam of 5~GeV, the lowest one
considered there).  With the 12~GeV energy, JLab will be able to measure
$F_{\pi}$ for the momentum transfers up to $\sim 6$~GeV$^{2}$ with the
accuracy of $\sim 4\%$. For EIC, there exist various proposals under
consideration; for the 5~GeV proton energy, $F_{\pi}$ might be measured up
to $Q^{2}\sim 15$~GeV$^{2}$ with the accuracy of $\sim 10\%$.

\begin{figure}
\centering
\includegraphics[width=0.67\columnwidth]{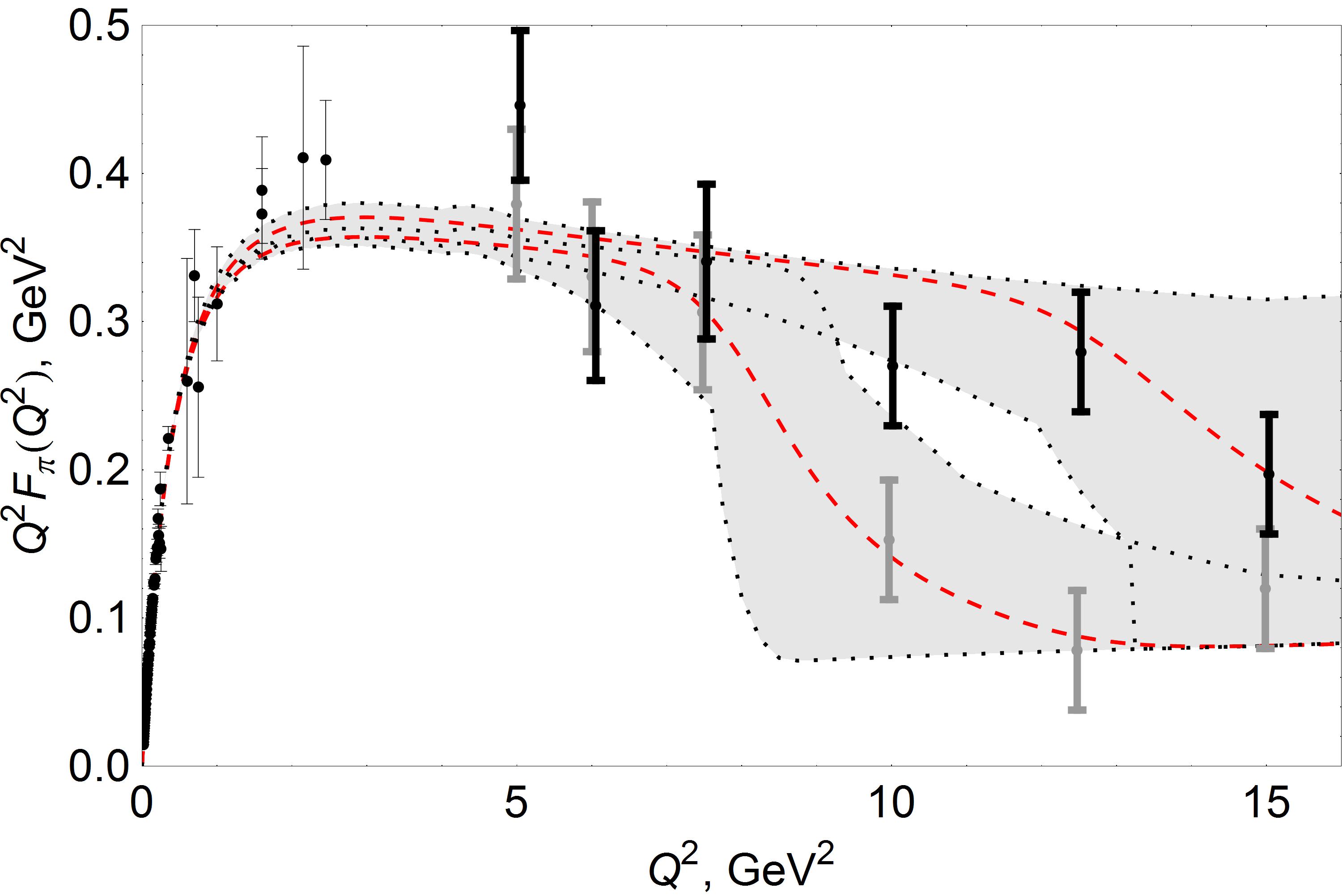}
\caption{
The range of predictions of our model for the charged pion form factor
allowed by simulated EIC data and model consistency for
two particular scenarios for $M(Q^{2})$ (dashed lines) corresponding to the
``low-scale'' and ``high-scale'' soft/hard transitions discussed in the
text (gray bands bound by dotted lines). Existing experimental data points
(see Ref.~\cite{PRD} for their description and list of references) are
shown by black dots with thin error bars.
Two typical simulated example realizations of expected EIC data,
corresponding to the ``high-scale'' (black) and ``low-scale'' (gray)
scenarios, are shown by dots with thick error bars.
JLab 12-GeV simulated data are not shown for clarity. It is evident that
the EIC data would make it possible to choose between the allowed scenarios
of quark-mass running within our model.}
\label{fig:plot-15GeV2}
\end{figure}

To proceed further, we restrict ourselves to two particular representative
scenarios corresponding to the ``low-scale'' and ``high-scale'' soft/hard
transition. The $(\mu, \lambda)$ parameters of these scenarios are shown
in Fig.~\ref{fig:mu-lambda} by symbols. The low-scale scenario corresponds
to the lowest transition scale which agrees, at the 95\% CL, with the
present data ($\mu=2.79$; $\lambda=0.715$). The high-scale one is a
typical representative point inside the allowed region ($\mu=3.5$;
$\lambda=1.0$). The corresponding $F_{\pi} (Q^{2})$ functions are shown in
Figs.~\ref{fig:plot-6GeV2}, \ref{fig:plot-15GeV2}
by dashed lines.

Having assumed particular values for $\mu$ and $\lambda$, and therefore a
particular $F_{\pi} (Q^{2})$ model curve, we simulate fake data points for
a given experiment, scattered around the theoretical curve with the
Gaussian distribution. The width of the distribution is determined by the
error bars quoted in Refs.~\cite{PR12-06-101, EIC-talk}; the values of
$Q^{2}$ for these fake ``measurements'' are also taken from there. Then,
these fake data are processed in the joint chi-square fit with the
existing data. The results are presented in Figs.~\ref{fig:mu-lambda},
\ref{fig:plot-6GeV2}, \ref{fig:plot-15GeV2}.

For 12-GeV JLab, one may see from Fig.~\ref{fig:plot-6GeV2} that the
expected error bars of the $F_{\pi}$ measurements exceed the width of the
region allowed by the present data for our model. Therefore, these data
are not expected to contribute much into the determination of the
soft/hard transition scale, as illustrated in Fig.~\ref{fig:mu-lambda}  in
terms of $(\mu,\lambda)$. However, we point out that the 12-GeV JLab data will
be of crucial importance for testing the model itself. In
Fig.~\ref{fig:plot-6GeV2}, predictions of two alternative scenarios
describing soft/hard transitions are also shown (see Ref.~\cite{PRD} for a
more detailed discussion). Clearly, the precision of the expected 12-GeV
JLab data would be sufficient to confirm or exclude the model we use here.

Contrary, the 5-GeV EIC data, though having large expected error bars,
will be able to disentangle the low-scale and high-scale transition
scenarios within our approach, see Figs.~\ref{fig:mu-lambda},
\ref{fig:plot-15GeV2}. The 95\% CL constraints obtained in
the two example scenarios are clearly separated. There still remains a
formal degeneracy for $Q^{2}_{\rm trans} \gtrsim 16$~GeV$^{2}$, but this
range of momentum transfer, with the energy scale of order the $b$-quark
mass, most probably corresponds to the perturbative QCD regime anyway.

\section{Conclusions}
\label{sec:concl}
We considered the effect of future experimental data on our understanding
of the dependence of the pion electromagnetic form factor, $F_{\pi}$,
on the momentum transfer squared, $Q^{2}$, paying a special attention to
the ``soft/hard'' transition region where the QCD asymptotics should
settle down. We took advantage of a particular low-energy model which
describes excellently the existing data and predicts the QCD asymptotics
automatically, without parameter tuning.

Given the estimated precision of 12-GeV JLab and of EIC, as well as the
expected range of the momentum transfer accessible to the instruments, we
conclude that the coming JLab data may confirm or exclude our model while
the EIC measurements would be able to pin down the soft/hard transition
scale.

\begin{acknowledgments}
The work of ST was supported
in part by the RFBR grant 12-02-01203.
\end{acknowledgments}

\end{document}